\documentstyle{article}

\newcommand{\nl}{ {\hfill \break} }
\newcommand{\np}{ {\newpage } }
\newcommand{\ol}{ \overline}

\newcommand{\iso}{ {\cong } }
\newcommand{\lto}{ {\longrightarrow } }
\newcommand{\imp}{ {\Rightarrow } }
\newcommand{\inc}{ {\hookrightarrow } }
\newcommand{\up}{ {\uparrow } }
\newcommand{\X}{ {\rm X} }
\newcommand{\Y}{ {\rm Y} }
\newcommand{\Z}{ {\rm Z} }
\newcommand{\eps}{ \mbox{$\varepsilon$} }
\newcommand{\cTP}{ \mbox{$T^*\rm P^1 $} }
\newcommand{\cTS}{ \mbox{$T^* S^1 $} }
\newcommand{\Po}{ \mbox{$\rm P^1 $} }
\newcommand{\s}{ \mbox{$ S^2 $} }
\newcommand{\R}{ \mbox{$\rm R $} }
\newcommand{\cd}{ \mbox{$\rm C^3 $} }
\newcommand{\cz}{ \mbox{$\rm C^2 $} }
\newcommand{\LT}{ \mbox{$\rm L_{+}^\up $} }

\newcommand{\Bri}{{1}}
\newcommand{\Du }{{2}}
\newcommand{\Hu }{{3}}
\newcommand{\Ki }{{4}}
\newcommand{\Kl }{{5}}
\newcommand{\Ur }{{6}}
\newcommand{\Gr }{{7}}
\newcommand{\Ra }{{8}}
\newcommand{\aEl}{{9}}
\newcommand{\Qa }{{10}}
\newcommand{\Sa }{{11}}
\newcommand{\Haw}{{12}}
\newcommand{\An }{{13}}
\newcommand{\Har}{{14}}
\newcommand{\Bra}{{15}}
\newcommand{\Pe }{{16}}
\newcommand{\Ar }{{17}}
\newcommand{\bEl}{{18}}
\newcommand{\Hi }{{19}}

\title{\Large {\bf
Resolution of simple singularities yielding \\
particle symmetries in a space-time } }
\author
{\bf{Martin Rainer}  \vspace{0.3truecm}\\
\normalsize\bf{Projektgruppe Kosmologie im Fachbereich Mathematik}\\
\normalsize\bf{Universit\"at Potsdam, Am Neuen Palais 10}\\
\normalsize\bf{D-O-1571 Potsdam, Germany}}
\date{}
\begin{document}
\large
\maketitle
\begin{abstract}
\large
\vspace{0.4truecm}
\hspace*{-0.75truecm}
A finite subgroup of the conformal group SL(2,C) can be related
to invariant polynomials on a hypersurface in ${\rm C}^3$.
The latter then carries a simple singularity, which resolves by a
finite iteration of basic cycles of deprojections.
The homological intersection graph of this cycles is the
Dynkin graph of an ADE Lie group. The deformation of the simple
singularity corresponds to ADE symmetry breaking.

A $3+1$-dimensional topological model of observation is constructed,
transforming consistently under SL(2,C), as an evolving $3$-dimensional
system of world tubes, which connect ``possible points of observation".
The existence of an initial singularity for the 4-dimensional space-time is
related to its global topological structure.

Associating the geometry  of ADE singularities to the vertex structure of
the topological model puts forward the conjecture on a likewise
relation of inner symmetries of elementary particles to local space-time
structure.

\vspace{1.1truecm}

\vspace{0.35truecm}
AMS Nr.: 20D06, 32S30, 32S45, 81T30, 83C75, 83F05
\end{abstract}

\vspace{0.4truecm}

\section{\bf Introduction}
\setcounter{equation}{0}
Relations between finite simple groups, simple singularities and simply
laced Lie groups, denoted by the Cartan series A, D and E,
are wellknown$^{\Bri-\Ur}$, but have not been used so far
to tie up a particle's inner symmetry, expressed by an ADE Lie group,
explicitly to local space-time geometry near the particle.
This geometry is assumed to be given as a vertex structure in a
$3+1$-dimensional topological model of observation,
which is based on a network of spatial world tubes evolving in time.
The present approach, inspired by string theory$^\Gr$, differs
from that by working a priori in $3+1$ dimensions. Breaking of
conformal symmetry is considered here as naturally given by the global
topology of space-time and the structure of observation.
The following section defines the topological model and discusses
democracy between ``possible points of observation" (PPOs) in its
relation to a possible initial singularity of the space-time and its global
topology.

Sec. 3 presents an overview over the conformal action of SL(2,C) and its
relatives on the structures of the model.

Sec. 4 resumes known results$^{\Bri-\Kl}$ on the finite subgroups
of SL(2,C), associated simple singularities, their resolutions and
homological relation of the latter to the ADE Lie groups.

Sect. 5 describes the singularities of the A and D series explicitly on
characteristic algebraic hypersurfaces in $C^3$, following essentially
the original work of F. Klein$^\Kl$.

Sec. 6 reviews deformations of simple
singularities$^{\Du,\Hu,\Ur}$ and presents a theorem$^\Ur$,
wellknown to algebraic topologists rather than physicists,
which allows to decide
the existence of certain deformations of the singularities by a graph
theoretical method. By the latter the relation to symmetry
breaking transitions between the associated ADE Lie groups becomes evident.
This together with Sec. 3 has recently encouraged  the author to
conjecture$^\Ra$ a close relation of ADE symmetries and local structure of
observation in space-time.

In Sec. 7 the geometric structure of the resolution of
singularities of series A is described explicitly. A properly chosen real
section can be identified  with some vertex of the model presented in
Sec. 2.
This confirms the conjecture above.

Finally we discuss the perspective of results for this model.

\section{\bf Topological model of space-time observation}
\setcounter{equation}{0}
Starting with $n+1$ vertices and demanding
any of them being connected to any other one by a stringy world tube,
the lowest dimension allowing the connections to be nonintersecting
is $d=3$, which shall be the case. Up to Sec. 6 the vertices are assumed
to contain a ``possible point of observation" (PPO) each.
{}From any vertex there are $n$ tubes going out, as the
only possibility to exchange information with other vertices.
The whole worldsheet shall be closed.
This structure is assumed to be embedded in some spacelike section $M^3$ of
the Lorentzian part $M^4\cong M^3\times \R$ of a
classical space-time and to evolve in $M^4$ according to a synchronized
time of one of the PPOs, which will be called ``the observer" O.
Since all $n+1$ PPOs are equal, anyone of them could
observe a likewise system of $n$ vertices. This fact is called ``democracy".
Fig. 1 shows an example of 5 vertices connected to each
other and all other ones.

Unification/separation of a single pair of vertices yields
separation/ unification of $n$ pairs of points on the worldsheet
after pinching/before blowing up n (closed) strings respectively:
\medskip \nl
{\normalsize
\begin{tabular}{ll} \quad
\begin{tabular}{llll}
{unify 2 PPOs}& $\imp$& n-times pinch \& split &{$ S^1 \lto \cdot \lto : $}
\\
{split 1 PPO} & $\imp$& n-times glue \&blow up
&{${\,}:{\,}\lto\cdot\lto S^1 $}
\end{tabular} &\quad  (2.1)
\end{tabular} }
In (2.1) the topological transformations of pinching
or blowing up of a closed string $S^1$ are denoted resp. by
$S^1\lto \cdot$ and $\cdot\lto S^1$,
while $\cdot \lto :$ denotes the separation of a pair of points $(:)$ by
splitting a single point $(\cdot)$ and $: \lto \cdot$ is the inverse
operation, the unification of the pair, gluing it into a single point.
\vspace{5.0truecm}
\begin{center}
{\normalsize Fig. 1: Net of vertices of PPOs connected by stringy tubes}
\end{center}
After O has been chosen and the
connecting tubes between all observed vertices removed, the worldsheet
connected to the observer can be deformed smoothly into a sphere
\s, centered at O, with $n$ punctures, corresponding to the observed
vertices.
This can be achieved by conformal rescaling of the original worldsheet
metric $h_0$ to $h=e^{\Phi} h_0$ by an appropriate choice$^\Gr$ of some
scalar field $\Phi$ on the tubular world sheet. The removed
connections between the observed vertices would correspond to
$n(n-1)/2$ handles attached to \mbox{\s.}

In a continuum limit $n \to \infty$ the punctures on \s are assumed to
become dense on the whole \s. Assuming a homogeneous distribution
of punctures implies local rotation symmetry$^\aEl$ on $M^3$.

Since O is at the center and all others PPOs are on \s, one might  think
that the original democracy is broken by observation. Whether this is so
depends however on the global topology of $M^3$. \nl
If e.g. $M^3\cong S^3$ (Fig. 2 a),
it is possible to embed O together with \s in $M^3$.
By shrinking \s , O can be exchanged with an other PPO
by an arbitrarily small deformation on $M^3$. Such a
situation is given in Bianchi IX models, e.g. de Sitter space-time,
which is free of an initial singularity. \nl
If on the other hand e.g. $M^3\cong \s\times \R$ (Fig. 2 b),
embedding \s symmetrically in $M^3$ excludes O from $M^3$.
Thus it cannot be exchanged with any other PPO.
Such a situation appears in Kantowsky Sachs (KS) models,
which have an essential singularity (i.e. one which cannot be removed
by coordinate transformations) at a point or even an interval of
time$^\Qa$.
\vspace{5.0truecm}
\begin{center}
{\normalsize Fig. 2: Democracy a) conserved in $M^3\cong S^3$,
b) broken in $M^3\cong\s\times \R$}
\end{center}
Singularities of spherical space-times like those mentioned above have been
classified recently in Ref. \Qa. One is led
to the conjecture that singularities in space-time are essentially effected
by a breaking of the permutation symmetry between all PPOs (vertices).

Note that the affliction of classical space-times by singularities mentioned
above refered to the case of Einstein-Hilbert (EH) theory, given by a
Lagrangian $L\!=\!R$.
For theories with additional higher order curvature invariants in $L$
a singularity which appears for a certain model in EH theory may be
evitable$^\Sa$, since the strong energy condition may not be valid and then
the Hawking-Penrose  theorem$^\Haw$ does not apply.

A resolution of scale factor singularities$^\Qa$ of the
classical space-time in the quantum model$^{\An,\Har}$
corresponds$^\Ra$ to blowing up O to $S^3$.
Note that quantum theory even more than
special relativity (SR) does not admit a separation of O from the
space-time.
The usual projection \s $\to$ C used to evaluate string theory as an
effective QFT in flat 2-space yields a conformal anomaly$^\Gr$ for
dimensions $D \ne 26$. Note that this projection simultaneously destroys
the symmetry between O and other PPOs.

Finally the alternative approach of implementing the principle of
SR at the quantum level by using Quantum Frames of Reference (QFR),
yields a quantized space-time$^\Bra$. This should be consistent with
the present multi-vertex approach by attaching a QFR to each vertex.

\section{\bf Action of SL(2,C) on the model structures}
\setcounter{equation}{0}
Consider now the group of invertible complex $2\times 2$ matrices GL(2,C).
The elements of determinant 1 form a normal subgroup SL(2,C). Division
of GL(2,C) resp. SL(2,C) by its center $\rm C^*$ resp. $\{\pm E\}$
yields the group PGL(2,C) resp. PSL(2,C), which is the analogue in
projective space, and can be identified with the automorphisms of \s
(or the M\"obius tranformations Aut $\rm\hat C$) resp. the proper
ortochronous Lorentz group $\rm L_{+}^\uparrow$
(or the 3-dim. complex orthogonal group). One gets the
following diagram with exact vertical sequences and an exact horizontal
splitting one:\nl
\medskip
{\normalsize
\begin{tabular}{cccccccccl}
&& &1& &1& & & & \\
&& &$\up$& &$\up$& & & & \\
SO(3,C)&=&$\rm L_{+}^\up =$&PSL(2,C)&&PGL(2,C)&= Aut $\rm\hat C$&=&Aut\s&\\
&& &$\up$& &$\up$& & & & \\
&1&$\inc$&SL(2,C)&$\to$&GL(2,C)&
${{ det\atop\lto} \atop \hookleftarrow}$&$\rm C^* $&$ \to\ 1$&  \\
&& &$\up$& &$\up$& & & & \\
&& &$\{\pm E\}$& &$\rm C^*$& & & & \\
&& &$\up$& &$\up$& & & & \\
&& &1& &1& & & (3.1)&
\end{tabular}
\medskip }
By (3.1) {\it transformations of \cz-spinors}, SL(2,C),
project to {\it transformations of bivectors}, SO(3,C),
and {\it proper orthochronous Lorentz transformations}, \LT,
and they essentially constitute (modulo dilatations and
phase factors){\it conformal transformations}, PGL(2,C),
on the Riemann sphere or
projective plane, which correspond conformally to transformations of a
tubular worldsheet$^\Gr$. Thus the transformation properties of the model of
Sec. 2 are fixed, since all its structures are covered by the
transformations above. In this paper only the worldsheet as observational
structure and the 4-dimensional space-time are considered.
(For a transcription to the language of spinors and their relation to the
light cone see Ref. \Pe.)

In the following section we will deal with another
very important structural property of SL(2,C), namely its finite subgroups,
which are related uniquely to the simple singularities.

\section{\bf SL(2,C) subgroups and simple singularities}
The finite subgroups F of SL(2,C), together with the properties of the
related simple singularities$^{\Bri,\Du,\Ki}$, are listed
in the following table:
\medskip\nl
{\normalsize
\begin{tabular}{llrrcl}
F    & name           &order&$R$(X,Y,Z) \qquad &{\ }basic graph\qquad& Lie G
\medskip\\
$C_n$&cyclic      &$n${\ } &$\X^n +{\ }\Y^2{\,}+\Z^2  $&& ${\rm A}_{n-1} $\\
$D_n$&binary dihedral &4$n${\ } &$\X^{n+1}\!+\!\X\Y^2\!+\Z^2$&&
${\rm D}_{n+2}$\\
$T$&binary tetrahedral&24{\ } &$\rm X^4 +{\ }Y^3{\,}+Z^2  $&& $\rm E_6  $\\
$O  $&binary octahedral &48 {\ }&$\rm X^3Y +{\ }Y^3{\,}+Z^2 $&& $\rm E_7$\\
$J  $&binary icosahedral&120{\ } &$\rm X^5  +{\ }Y^3{\,}+Z^2 $&& $\rm E_8$
\end{tabular}
\begin{center}
Table: Properties of finite SL(2,C)-subgroups and simple singularities
\end{center}
}\hspace*{-0.75truecm}
The A and D series are running over integers $n\ge 2$.
Any F is
located in SU(2), the maximal compact subgroup of SL(2,C) and the double
cover of SO(3,R). That yields (for F$\neq C_{2j+1}$) the fibration
\begin{center}
{\normalsize
\begin{tabular}{cr}\qquad\quad
\begin{tabular}{ccccccccc}
1& $ \to $ &$\{\pm E\}$ & $\inc$ & SL(2,C) & $\to$ & SO(3,C) &$ \to $&1\\
&&  & & $\cup$ & & $\cup$ && \\
1& $ \to $ & $\{\pm E\}$ &$\inc $&SU(2) &$\to$ & SO(3,R) &$ \to $&1 \\
&&  & & $\cup$ & & $\cup$ && \\
1& $ \to $ &$\{\pm E\}$ & $\inc$ & F & $ \to $& PF &$ \to $&1
\end{tabular} & \qquad  (4.1)
\end{tabular} }
\end{center}
\medskip
(4.1) explains the names of the F, since PF is the symmetry group of a
regular polyhedron.

Let us now consider the  F-invariant polynomials on $\rm C^2$.
They are generated by 3 fundamental F-invariants  X, Y, Z  subject to 1
constraint  {$R$(X,Y,Z)=0}. The normal forms of  $R$(X,Y,Z) are listed in
the table above. Thus we have a quotient map
$$ \qquad \qquad \quad \qquad
q :\left\{ \begin{array}{ccc}
		\rm C^2 & \to &\rm C^3 \\
   {z_1 \choose z_2}\equiv z & \mapsto & \left( \begin{array}{c}
						{\rm X} ( z )\\
						{\rm Y} ( z )\\
						{\rm Z} ( z )
					       \end{array} \right)
	      \end{array} \right. \qquad \qquad \qquad (4.2)
$$
The image $q(\rm C^2) = C^2 / F$ is a hypersurface in $\rm C^3$ given by
the constraint R. For any F a simple singularity (at 0) is just given by
(the complex analytic germ of)
$\rm C^2 / F$. For a complete classification list of all hypersurface
singularities see Ref. \Ar. Here we restrict to the simple ones.
Their properties have already been examined by F. Klein$^\Kl$. A modern
overview is given by Ref. \Du.

In the following we need the notion of a resolution of a singular variety
$X$ (here $X = \rm C^2 / F$). This is given by a surjective proper map
$\pi : Y \to X$
from a regular variety $Y$ to the singular $X$, such that the regular subset
$ X_{reg}$ (=$X\!-0$ in the present case) is diffeomorphic
with its pre-image $\pi^{-1} \ X_{reg}$, which is required to be dense
in $Y$.
Normally we want $Y$ to be a minimal resolution, demanding
$\forall \pi',Y'\ \exists j : Y \inc Y'$ such that $ \pi' \circ  j = \pi$.

It is now a very special property of the simple singularities that they can
be resolved by a finite iteration of a deprojection proceedure called
blow up or $\sigma$-process$^{\Bri,\Ki}$.
For $\rm C^3$ the blow up at 0 is given by the canonical projection
$\beta$ of the tautological line bundle
$\tau  = \{(x,l) \in {\rm C^3 \times P^2} : x \in l \}$
onto $\rm C^3$. The singular fiber of $\beta$ is given by the complex
projective space $\rm P^2$.
Then the blow up $\hat X$ of a subvariety $X \subset \cd$ at 0 is defined
by the closure of the pre-image of $X_{reg}=X-0$ under $\beta$
$$       \qquad \qquad     \qquad \qquad
\begin{array}{ccl}
\cd & {\beta\atop\gets} & \tau \\
\cup & &\cup \\
X &\gets&\hat X = \ol{\beta^{-1}(X-0)} \\
\cup & &\cup \ \mbox{\small\it dense} \\
X-0 &\gets&\beta^{-1}(X-0) \\
\end{array}  \qquad \qquad \qquad (4.3)
$$
Assume $\pi = \beta^n : Y \to X$ is a resolution of a simple singularity.
The exeptional set, defined as
$E = { Y} / \pi^{-1} X_{reg}$,
is a finite union of
homological cycles $\rm C_1, ..., C_r$ with each $\rm C_k$ isomorphic to
${\rm P^1} \iso \frac{S^3}{S^1} \iso S^2$. So the cycles are actually
2-spheres. The first Chern class of the normal bundle of $\rm C_k$ in $Y$ is
${\rm c_1} ({\rm C_k},Y) = -2$,
which is also known as the selfintersection number$^\Hu$ of $\rm C_k$ in $Y$.

In fact the cycles $\rm C_k$ here either intersect in 1 point or do not
intersect at all. So we can assign a basic graph to a simple singularity
representing each cycle of its resolution as a dot and connecting 2 dots
by an edge if the cycles intersect. It turns out that any basic graph of a
simple singularity corresponds to the Dynkin graph of a simply laced
Lie group.
The last two columns of the table above show the basic graph and the
corresponding Lie group in Cartan's notation.

\section{\bf Explicit Description of the A and D Series}
\setcounter{equation}{0}
Let us now consider some instructive examples of the
singularities explicitly. First the A series: \nl
The constraint $ R = \X^n + \Y^2 + \Z^2 = 0$ for the
cyclic singularities
is equivalent to
$Q = \X^n+\rm U V = 0$ by the substitutions
\begin{center}
\begin{tabular}{l}
U = {Y + $i$ Z} \vspace{0.15cm}\\
V = {Y $-$ $i$ Z} \\
\end{tabular}
respectively
\begin{tabular}{l}
Y = {${1\over 2}\,$}\,{(U + V)} \vspace{0.15cm}\\
Z = {${1\over 2i}$}\,{(U $-$ V)} \\
\end{tabular}
\end{center}
In SL(2,C) the cyclic group of order $n$ is represented as \nl
$C_n=\{1,E_n,..,E_n^{n-1}\}$ with
$$
E_n = \left(
\begin{array}{cc}
e^{i\frac{2\pi}{n} } & 0       \\
  0                & e^{-i\frac{2\pi}{n} }
\end{array}
\right).
$$
Obviously X=\,$z_1 z_2$, U=\,$i z_1^n$, V=\,$i z_2^n$ are 3 independent
$C_n$-invariants, satisfying Q=0. Thus the 3 generic
$C_n$-invariant polynomials with R=0 are
$$ \qquad \qquad \qquad\qquad\qquad
\begin{array}{l}
\X = z_1 z_2 \vspace{0.15cm}\\
\Y = \frac{i}{2}\, (z_1^n + z_2^n) \vspace{0.15cm}\\
\Z = \frac{1}{2}\, (z_1^n - z_2^n)
\end{array} \qquad\qquad\qquad\qquad \quad (5.1)
$$
Now let us go over to the D series:\nl
The binary dihedral group $D_n$ is represented in SL(2,C) by combining
$C_{2n}$ with
$$
B = \left(
\begin{array}{cc}
0 & 1 \\
  -1  & 0
\end{array}
\right).
$$
Note that $B^2 = E_{2n}^n = \{-E\}\in C_{2n}$.
It is now easy to verify that
$$ \qquad \qquad\qquad\qquad\quad
\begin{array}{l}
\X = (z_1 z_2)^2 \vspace{0.15cm}\\
\Y = \frac{i}{2}\, (z_1^{2n} + z_2^{2n}) \vspace{0.15cm}\\
\Z = \frac{1}{2}\, z_1 z_2 \,(z_1^{2n} - z_2^{2n})
\end{array}  \qquad\qquad\qquad \qquad (5.2)
$$
are 3 independent $D_n$-invariant polynomials, which satisfy just the
constraint $ R = \X^{n+1} + \X\Y^2 + \Z^2 = 0$ given above. \nl
The lowest order nontrivial group of this series is $D_2$.
Its projection onto
$L_+^{\up}$ yields ${\rm P}D_2 = V_4 \equiv \{E,C,P,T\}$, the
Kleinian 4-group, which is the unique abelian group of order 4 with
$C^2=P^2=T^2=E$ and $CPT=E$.

Similar like the invariants (5.1) and (5.2) for the A resp. D series
one can construct the invariants of type E.
The results are already contained in the original work$^\Kl$ of F. Klein.
The following section is devoted to the behavior of the ADE singularities
under algebraic perturbations.

\section{\bf Deformations and symmetry breaking}
\setcounter{equation}{0}
For the singularity given by $R$(X,Y,Z) = 0
let us define (following Refs. \Du, \Hu\ and \Ur)
small deformation (fiber)s
in an $\eps$-neighbourhood of 0 by
$$
M_{S,t}\!=\!\{(\X,\Y,\Z)\!\in\!B^3(\eps)\!\subset\!\cd \mid
	      R(\X,\Y,\Z) + t S(\X,\Y,\Z)\!=\!0 \} \quad (6.1)
$$
with $t$ small w.r.t.\eps.\nl
Clearly $M_{S,0}$ is the singularity itself. $M_{1,t}$ is the Milnor fiber,
which characterizes the resolution, since $H_2(M_{1,t},\Z)$ is just
generated by the cycles $\rm C_k$ introduced in sec. 4.
Following theorem$^{\Ur}$ holds:\nl
{\it
Given a simple singularity and independently a Dynkin graph G of ADE type
(i.e.\,each  connected component of G is of ADE type).
\nl
Then the following properties are equivalent:
\nl
I. There exists a deformation $M_{S,t}$ with only simple singularities,
\hspace*{0.6cm} such that the combination of these corresponds exactly to G.
\nl
II. G is a subgraph of the basic graph of the singularity.
}

Similar theorems hold also for larger classes of singularities$^\Ur$.
The powerful implication of the theorem above is that it allows to decide
just graph theoretically the existence of both, deformations of simple
singularities and corresponding symmetry breaking transitions$^\Ra$
between the associated ADE Lie groups (see example in Fig. 3).
\vspace{3.5truecm}
\begin{center}
{\normalsize Fig. 3: Example of possible symmetry breaking}
\end{center}
Considering also the actions of
SL(2,C) on space-time and tubular worldsheet according to Sec. 3,
the author has conjectured$^\Ra$ an intimate relation between space-time and
observational structure on the on hand and internal (Yang Mills like)
symmetries on the other.
\nl

In the following section we describe the geometric structure of some
simple singularities explicitly and relate them to the structure of the
model of Sec. 2, thus confirming the conjecture above.

\section{\bf Geometry of the singularities in the model}
\setcounter{equation}{0}
Let us consider the geometry of the most simple singularity
$\rm A_1$ given by the constraint
$$    \qquad  \qquad \qquad\qquad\qquad
- \X^2 + \underbrace{(i \Y)^2 + (i \Z)^2}_{R^2} = 0 .
\qquad \qquad   \qquad \ (7.1)
$$
This surface is a double cone with double point singularity at 0,
the 2 cones corresponding to $X=\pm R$. It is resolved by a single
blow up (4.3)
yielding a surface isomorphic to $T^*\rm P^1$ and the singular fiber is
given by the zerosection $P^1$ (see Fig. 4).
$T^*\rm P^1$ is the unique line bundle of Chern class -2 and thus it is
the standard building block for the resolutions of higher ADE singularities
involving more fundamental cycles.
\vspace{8.0truecm}
\begin{center}
{\normalsize Fig. 4: Resolution of $A_1$}
\end{center}
\nl
More generally an ${\rm A}_n$ singularity is given by a surface
$\X^{n+1}=R^2$,
with $R=i(\Y^2+\Z^2)^{\frac{1}{2}}$ like above.
This surface is built by $n+1$ cones given by
$\X=e^{\frac{2\pi i k}{n+1}} R^{\frac{2}{n+1}}$ with $k=1,..,n+1$,
their tops joining in the singularity 0. The
resolution can be found by $n$ blow ups (4.3), with $n-i$ cones left after
the $i$th blow up. The resolution is given by a vertex of $n+1$ joining
tubes \cTP rather than cones. The singular fiber of the resolution
is given by a joining chain of $n$ spheres $\Po\vee \Po..\vee \Po$.

This is now demonstrated in detail for $n=2$ (see Fig. 5).
The surface is given by
3 cones joining in a point. The first
blow up yields an $\hat X$ like a $T^*\Po$ tube with a cone
attached to the image of the
zerosection \Po in a point $\hat 0$. This is however not yet a resolution,
since $\ol{\beta^{-1}  X_{reg}}$ is singular in $\hat 0$.
Applying (4.3) once more
$\hat 0$ is blown up to a further \Po intersecting the first one just in a
point. This yields the resolution,
given by 3 regularly joining \cTP tubes.
The singular fiber is the join $\Po\vee\Po$ of 2 spheres.\nl
\vspace*{8.8 truecm}
\begin{center}
{\normalsize Fig. 5: Resolution of $A_2$}
\end{center}
\nl
In order to relate these complex structures to the real structures of our
topological model, let us take a proper real section, such that $\Po \iso
S^2$ becomes restricted to $S^1$ and the complex cotangent bundle \cTP
restricts to the real cotangent bundle $\cTS\iso S^1\!\times\!\rm R^1$.
Thus we can identify the tubular strings between PPOs as real sections
of resolutions of $\rm A_1$ singularities, which themselves might have
been created each
by 2 points of the worldsheet identifying to a double point. So
$\rm A_1$ singularities and resolution appear naturally in processes
like (2.1).

Until now we have considered any single PPO as vertex with $n$ tubes
attached.
Note however that this point of view can be transformed into a different
one
by a smooth deformation of the worldsheet replacing any $n$-tube vertex
replaced by $n-k+1$ vertices with maximally $k$ external tubes
($3\le k\le n$).
Taking such a vertex as PPO, the associated 2-sphere would have only $k$
punctures. The most refined decomposition
is the tree level with $k=3$.
At this level any vertex corresponds to a real
section of an $\rm A_2$ resolution.
Generally at level $k$ any maximal vertex corresponds to an
${\rm A}_{k-1}$ resolution.

However at any refined level  democracy among vertices is broken, since no
longer each of them is connected to any other one. Democracy of PPOs can
be retained if we consider only a whole cluster of $n-k+1$ ``confined"
vertices, rather than a simple vertex, as PPO.

Note that the resolution by passing in (6.1) from $M_{1,0}$ to
$M_{1,t}$ with
$t\ne 0$ is similar like in QFT shifting particles off the singular light
cone by giving them a (small) nonzero mass. In fact the geometry of the
$\rm A_1$-resolution corresponds to a complex mass shell.

We had chosen real sections, like e.g.\,in (7.1), such that they
correspond to the
worldsheet structure in a syncronized spacelike section $M^3$.
Note however that the complex structure of the resolution contains various
possibilities for real sections and choices of signatures. Recently it was
pointed out, how a change of space-time signature may be posssible$^\bEl$
and interpreted$^\An$,
where instantons play a crucial role. A full space-time
analysis of the model of Sec. 2 provides an instanton interpretation$^\Hi$
of geometric structures related to ADE singularities.

\section{\bf Discussion}
\setcounter{equation}{0}
Using a specific model of observation with special emphasis to microscopic
democracy of PPOs (``possible points of observation") in space-time,
conjectures about an intimate relation
between inner symmetries of the Yang Mills type and local structure
of observation in space-time near a particle are confirmed.\nl
Explicitly the resolution of singularities of the A series appear naturally
as vertices of the tubular network which is transporting the information
between PPOs. The $\rm A_1$ singularity and its resolution (corresponding to
the SU(2) symmetry  of electroweak interaction) appears in any
unification/separation process of (PPO) vertices as pinch/blow up
of specific strings on the tubular worldsheet.

Abelian U(1) factors
are not considered in the ADE theory but might be represented
by (counter)clockwise twistings of the stringtubes around
their axis according to the polarisation of a lightwave propagating along
some of them.

The resolutions of the ${\rm A}_{k-1}$ singularities for $k\ge 3$
(corresponding to a SU($k$) Yang Mills symmetry) have real sections
isomorphic to a $k$-tube vertex. Confinement might be related to the
necessity to consider clusters rather than single vertices as
PPOs in order to restore democracy at the most refined level $k=3$.

We have coined the term PPO for an undetermined location somewhere in the
interior at a vertex of an extended observer carrying the frame of
reference. In contrast to a
quantization of space-time into disconnected parts$^\Bra$, our model
assumes
(at least in the initial phase considered here) all PPOs to be located
in a common connected region, namely the interior of the closed tubular
worldsheet. Regions connected only by stringtubes with
diameter very much below Plancklength $L_{Pl}$ should be considered as
quasidisconnected, which yields
the correspondence to the QFR (Quantum Frame of Reference) approach$^\Bra$.

String pinchings according to (2.1) can be performed equivalently around
interior and exterior of the tubular worldsheet w.r.t. the embedding space,
both yielding the same unification of vertex tubes.
Note the
resemblance of the present approach to knot theory and neural networks.
\np
Assuming a mean extension of the vertices
of only few orders of magnitude above $L_{Pl}$ and
a mean stringtube diameter of order $L_{Pl}$ or below, the PPOs
become sharp only for
$L_{Pl} \to 0$ in the macroscopic continuum limit
$n\to \infty$. In this limit
there should appear a connection to the gauge bundle framework.

By the theorem$^\Ur$ of Sec. 6, the Dynkin graphs describe
the possible ADE symmetry breaking transitions and deformations
of singularities. Notably in our model all this happens a priori
in a space-time of $D=4$ rather than decending from $D=26$.

For an evolutionary process there remains therefore the crucial question,
whether a classical space-time (initial) singularity appears.
This question
is closely related to the question of PPO democracy via the global
topology of $M^3$.
Furthermore a resolution
of a scalefactor singularity corresponds to a deprojection process,
similar to those involved in the ADE resolutions.
All that indicates a unified description of both the
singularities {\it in} and {\it of} space-time. This and a more detailed
evaluation of the physical predictions of the model will be topic of a
forthcoming paper.
\nl\nl\nl
{\Large {\bf Acknowledgements}}
\nl\nl
The author would like to thank the unknown referee for his useful comments
on the paper. Furthermore he is grateful to
H.-J. Schmidt for critical remarks on the
topic and thanks the Cosmology Group of the Mathematics Department at
Potsdam University for their hospitality.
\np\hspace*{-0.75truecm}

{\Large {\bf References}}
\nl\nl
$^\Bri$  E. Brieskorn, Math. Ann. 166 (1966) 76
\nl
$^\Du$  A. Durfee, Enseign. Math. 25 (1979) 131
\nl
$^\Hu$  S.M. Husein-Zade, Russian Math. Surveys 32:2 (1977) 23
\nl
$^\Ki$ D. Kirby, Proc.\,London Math.\,Soc.\,(3), 6\,(1956) 597,
{7\,(1957)\,1}
\nl
$^\Kl$ F. Klein, Vorlesungen \"uber das Ikosaeder und die Aufl\"osung der
\hspace*{0.12cm} Gleichungen vom f\"unften Grade (Teubner, Leipzig, 1884)
\nl
$^\Ur$ T. Urabe, ICTP Lecture Notes SMR.567/4 (1991)
\nl
$^\Gr$  M.B. Green, J.H. Schwarz and E. Witten, Superstring Theory,
\hspace*{0.12cm}        Cambridge Univ. Press (1987)
\nl
$^\Ra$ M. Rainer, Projective Geometry for Relativistic Quantum
\nl
\hspace*{0.12cm} Physics, preprint 92/05, Univ. Potsdam (1992)
\nl
$^\aEl$  G.F.R. Ellis, J. Math. Phys. 8 (1967) 1171
\nl
$^\Qa$  A. Qadir, J. Math. Phys. 33 (1992) 2262
\nl
$^\Sa$ A.D. Sacharov, Dokl. Akad. Nauk SSSR 177 (1967) 70, cf.
\nl
\hspace*{0.12cm} the review: H. Fuchs, U. Kasper, D.-E. Liebscher,
V. M\"uller,
\nl
\hspace*{0.12cm} H.-J. Schmidt, Fortschr. Phys. 36,6 (1988) 427
\nl
$^\Haw$  S.W. Hawking and G.F.R. Ellis, The Large Scale Structure of
\hspace*{0.12cm} Space-Time (Cambridge Univ. Press, Cambridge, 1973)
\nl
$^\An$ Y. Anini, Quantum Tunneling and the Change of Signature of
\nl
\hspace*{0.12cm}  the Spacetime Metric, ICTP preprint 2435/92 (1992)
\nl
$^\Har$  J.B. Hartle and S.W. Hawking, Phys. Rev. D 28 (1983) 2960
\nl
$^\Bra$  V.P. Bransky,
	Quantum Theory of Relativity as Quantum Clep-
\nl
\hspace*{0.12cm} sodynamics,
	Proc. London Philosophical Conference (1992) 72
\nl
$^\Pe$ R. Penrose, W. Rindler, Spinors and Spacetime, Cambridge Univ.
\hspace*{0.12cm} Press (1986)
\nl
$^\Ar$  V. Arnold, Invent. Math. 35 (1976) 87
\nl
$^\bEl$  G.F.R. Ellis, Change of Signature in Classical Relativity,
\nl
\hspace*{0.12cm}  preprint SISSA 170/91/A (1991)
\nl
$^\Hi$  N. J. Hitchin, Math. Proc. Camb. Phil. Soc. 85 (1979) 465
\np
Figure captions:
\begin{center}
{\normalsize Fig. 1: Net of vertices of PPOs connected by stringy tubes}
\end{center}
\begin{center}
{\normalsize Fig. 2: Democracy a) conserved in $M^3\cong S^3$,
b) broken in $M^3\cong\s\times \R$}
\end{center}
\begin{center}
{\normalsize Fig. 3: Example of possible symmetry breaking}
\end{center}
\begin{center}
{\normalsize Fig. 4: Resolution of $A_1$}
\end{center}
\begin{center}
{\normalsize Fig. 5: Resolution of $A_2$}
\end{center}
\nl
Table caption:
\begin{center}
{\normalsize Table: Properties of finite SL(2,C)-subgroups and simple
singularities}
\end{center}
\end{document}